\documentclass[12 pt,twoside,aps,prd,amsmath,amssymb,
tightenlines,showpacs,showkeys]{revtex4}
\usepackage{mathptmx}
\usepackage{bm}
\usepackage{graphicx}
{\hspace*{\fill}{\protect\small {\bf Bijan Saha}} \hspace*{\fill} }
{\hspace*{\fill} {\protect\small {\bf Non-minimally coupled
nonlinear spinor field in Bainchi type-I cosmology}} \hspace*{\fill}
} 
\usepackage[colorlinks]{hyperref}

\newcommand {\G}{\Gamma}

\newcommand {\bg}{\bar \gamma}

\newcommand {\bp}{\bar \psi}

\begin{document}
\title{Non-minimally coupled nonlinear spinor field in Bianchi type-I cosmology}
\author{Bijan Saha}
\affiliation{Laboratory of Information Technologies\\
Joint Institute for Nuclear Research, Dubna\\
141980 Dubna, Moscow region, Russia\\ and\\
Institute of Physical Research and Technologies\\
People's Friendship University of Russia\\
Moscow, Russia} \email{bijan@jinr.ru}
\homepage{http://spinor.bijansaha.ru}

\hskip 1 cm

\begin{abstract}
Within the scope of Bianchi type-$I$ cosmological model we have
studied the role of spinor field in the evolution of the Universe.
In doing so we have considered the case with non-minimal coupling.
It was found that the non-diagonal components of the energy-momentum
tensor of the spinor field, hence the restrictions on the space-time
geometry remain the same as in case of minimal coupling. Since in
this case the diagonal components of the energy-momentum tensor
differ, the evolution of the corresponding universe also differs.
For example, while a linear spinor field with non-minimal coupling
or nonlinear spinor field with minimal coupling give rise to open
universe, a nonlinear spinor field with non-minimal coupling with
the same parameters can generate close universe that at the
beginning expands, and after attaining some maximum value begin to
contract and finally ends in a Big Crunch.
\end{abstract}



\maketitle

\bigskip

\section{Introduction}

For more than two decades spinor field is being widely used in
cosmology mainly thanks to its specific behavior in presence of
gravitational field. In a number of papers the authors have shown
that the nonlinear spinor field can give rise to regular solutions
as well as explain the late-time accelerated mode of expansion of
the Universe
\cite{Saha2001PRD,Saha2006PRD,Saha2009aECAA,ELKO,kremer,Saha2018ECAA}.
But most of those papers considered the non-minimal coupling of
spinor and gravitational field. Recently, Carloni {\it et al}
\cite{Astro-Phys/1811.10300} has considered non-minimally coupled
spinor field with the gravitational one. In this report we plan to
generalize our earlier results for the interacting gravitational and
spinor fields.

\section{Basic equations}

We consider the action in the form

\begin{equation}
S = \int \sqrt{-g} \left[\left(1 + \lambda_1 S\right) R + L_{\rm sp}
\right] d \Omega. \label{action}
\end{equation}
where $S = \bp\psi$ is a scalar constructed from spinor fields,
$\lambda_1$ is the coupling constant. Let us work in natural unit
setting  speed of light $c = 1$ and Einstein's constant $\kappa =
1$. The spinor field Lagrangian takes the form

\begin{equation}
L_{\rm sp} = \frac{\imath}{2} \left[\bp \gamma^{\mu} \nabla_{\mu}
\psi- \nabla_{\mu} \bar \psi \gamma^{\mu} \psi \right] - m \bp \psi
- \lambda F (S). \label{lspin}
\end{equation}
Note that in general the nonlinear term $F$ may be the arbitrary
function of invariant $K$ which takes one of the following
expressions: $\{I,\,J,\,I+J,\,I-J\}$. Here $I = \bp \psi$ and $J =
\imath \bp \bg^5 \psi$. Here $m$ is the spinor mass. $\lambda$ is
the self coupling constant that can be positive or negative. Here
$\nabla_{\mu}$ is the covariant derivative of the spinor field so
that
\begin{equation}
\nabla_\mu \psi = \partial_\mu \psi - \Gamma_\mu \psi, \quad
\nabla_\mu \bp = \partial_\mu \bp + \bp \Gamma_\mu. \label{covder}
\end{equation}
Here $\Gamma_\mu$ is the spinor affine connection which can be
defined as
\begin{eqnarray}
\Gamma_\mu &=& \frac{1}{8}\left[\partial_\mu \gamma_\alpha,
\gamma^\alpha\right] - \frac{1}{8}
\Gamma^{\beta}_{\mu\alpha}\left[\gamma_\beta, \gamma^\alpha\right].
\label{SPAC}
\end{eqnarray}
where $\left[a,b\right] = a b - b a.$ Here $\gamma_\beta =
e_\beta^{(b)} \bg_b$ and $\gamma^\alpha = e^\alpha_{(a)} \bg^a$ are
the Dirac matrices in curve space-time and $e^\alpha_{(a)}$ and
$e_\beta^{(b)}$ are the tetrad vectors. The $\gamma$ matrices obey
the following anti-commutation rules
\begin{eqnarray}
\gamma_\mu \gamma_\nu + \gamma_\nu \gamma_\mu = 2 g_{\mu\nu}, \quad
\gamma^\mu \gamma^\nu + \gamma^\nu \gamma^\mu = 2 g^{\mu\nu}.
\nonumber
\end{eqnarray}
Variation with respect to metric functions give
\begin{equation}
R_{\mu\nu} - \frac{1}{2} g_{\mu\nu} R = \frac{1}{\left(1 + \lambda_1
S \right)}\left[T_{\mu\nu} + \lambda_1 \left(\nabla_\mu \nabla_\nu -
g_{\mu\nu} \Box\right)S\right]. \label{EE1}
\end{equation}
In our case it will be convenient to write the forgoing equation in
the following way
\begin{equation}
R_\mu^\nu - \frac{1}{2} \delta_\mu^\nu R = \frac{1}{\left(1 +
\lambda_1 S \right)}\left[T_\mu^\nu + \lambda_1
\left(g^{\nu\tau}\nabla_\mu \nabla_\tau - \delta_\mu^\nu
\Box\right)S\right]. \label{EE}
\end{equation}
where $T_\mu^\nu$ is the energy-momentum tensor of the spinor field.
The corresponding equations for spinor field we find varying the
action with respect to $\psi$ and $\bp$. In this case we find

\begin{subequations}
\label{speq}
\begin{eqnarray}
\imath\gamma^\mu \nabla_\mu \psi - m \psi - \lambda F_S \psi + \lambda_1 R \psi &=&0, \label{speq1} \\
\imath \nabla_\mu \bp \gamma^\mu +  m \bp + \lambda F_S \bp  -
\lambda_1 R \bp &=& 0. \label{speq2}
\end{eqnarray}
\end{subequations}
From \eqref{speq} one finds that $L_{\rm sp} = SF_S - F.$ Let us
also note that though the covariant derivative acts on the spinor
field in accordance with \eqref{covder}, it acts on $S = \bp \psi$
just like that on a scalar field. Then taking into account that
$\nabla_\nu S = \partial_\nu S$, we find
\begin{subequations}
\label{nablaS}
\begin{eqnarray}
\nabla_\mu \nabla_\nu S &=& \nabla_\mu \partial_\nu S = \partial_\mu
\partial_\nu S - \Gamma^\alpha_{\mu \nu} \partial_\alpha S.
\label{nabla1}\\
\Box S &=& g^{\alpha \beta} \nabla_\alpha \nabla_\beta S = g^{\alpha
\beta} \left(\partial_\alpha
\partial_\beta S - \Gamma^\tau_{\alpha \beta} \partial_\tau
S\right). \label{nablaS2}
\end{eqnarray}
\end{subequations}
Let us now introduce the Bianchi type-I space-time

A Bianchi type-$I$ anisotropic space-time is given by
\begin{equation}
ds^2 = dt^2 - a_1^{2} \,dx_1^{2} - a_2^{2} \,dx_2^{2} -
a_3^{2}\,dx_3^2, \label{bi}
\end{equation}
with $a_1,\,a_2$ and $a_3$ being the functions of time only. It is
the simplest anisotropic model of space-time. The reason for
considering anisotropic model lays on the fact that though an
isotropic $FRW$ model describes the present day Universe with great
accuracy, there are both theoretical arguments and observational
data suggesting the existence of an anisotropic phase in the remote
past.

For the metric \eqref{bi} we choose the tetrad as follows:

\begin{equation}
e_0^{(0)} = 1, \quad e_1^{(1)} = a_1, \quad e_2^{(2)} = a_2, \quad
e_3^{(3)} = a_3. \label{tetradbi}
\end{equation}

From the \eqref{SPAC} one finds the following expressions for spinor
affine connections:
\begin{equation}
\G_0 = 0, \quad \G_1 = \frac{\dot a_1}{2} \bg^1 \bg^0, \quad \G_2 =
\frac{\dot a_2}{2} \bg^2 \bg^0, \quad \G_3 = \frac{\dot a_3}{2}
\bg^3 \bg^0. \label{sac123bi}
\end{equation}
We consider the case when the spinor field depends on $t$ only. The
spinor field equations in this case read

\begin{subequations}
\label{speqbi}
\begin{eqnarray}
\imath\gamma^0 \dot \psi + \imath \frac{\dot V}{V} \psi
- m \psi - \lambda F_S \psi + \lambda_1 R \psi &=&0, \label{speq1bi} \\
\imath \dot \bp \gamma^0 + \imath \frac{\dot V}{V} \bp + m \bp +
\lambda F_S \bp  - \lambda_1 R \bp &=& 0, \label{speq2bi}
\end{eqnarray}
\end{subequations}
where we define the volume scale $V = a_1 a_2 a_3$.

From \eqref{speqbi} one easily finds
\begin{equation}
S = \frac{C_0}{V}, \quad C_0 = {\rm Const.} \label{SV}
\end{equation}

The energy-momentum tensor of the spinor field
\begin{equation}
T_{\mu}^{\,\,\,\rho}=\frac{\imath}{4} g^{\rho\nu} \biggl(\bp
\gamma_\mu \nabla_\nu \psi + \bp \gamma_\nu \nabla_\mu \psi -
\nabla_\mu \bar \psi \gamma_\nu \psi - \nabla_\nu \bp \gamma_\mu
\psi \biggr) \,- \delta_{\mu}^{\rho} L_{\rm sp}. \label{temsp}
\end{equation}
on account of \eqref{covder} can be written as
\begin{eqnarray}
T_{\mu}^{\,\,\,\rho}&=&\frac{\imath}{4} g^{\rho\nu} \biggl(\bp
\gamma_\mu
\partial_\nu \psi + \bp \gamma_\nu \partial_\mu \psi -
\partial_\mu \bar \psi \gamma_\nu \psi - \partial_\nu \bp \gamma_\mu
\psi \biggr)\nonumber\\
& - &\frac{\imath}{4} g^{\rho\nu} \bp \biggl(\gamma_\mu \G_\nu +
\G_\nu \gamma_\mu + \gamma_\nu \G_\mu + \G_\mu \gamma_\nu\biggr)\psi
 \,- \delta_{\mu}^{\rho} L_{\rm sp}, \label{temsp0}\\
 &=&  g^{\rho\nu} {\bar T}_{\nu\mu} -  g^{\rho\nu}
 {\tilde T}_{\nu\mu} - \delta_{\mu}^{\rho} (S F_S - F(S)), \nonumber
\end{eqnarray}

The nontrivial components of the energy-momentum tensor in this case
takes the form From \eqref{temsp0} for the nontrivial components of
the energy momentum tensor one finds \cite{Saha2006IJTP}:
\begin{subequations}
\label{Ttotbi}
\begin{eqnarray}
T_0^0 & = &  m S + \lambda F(S),\label{Ttot00bi}\\
T_1^1 & = & T_2^2 = T_3^3 = \lambda\left( F(S) - S F_S\right),\label{Ttotiibi}\\
T_2^1 & = & -\frac{\imath}{4} \frac{a_2}{a_1} \left(\frac{\dot
a_1}{a_1} - \frac{\dot a_2}{a_2}\right)  \bp \bg^1 \bg^2 \bg^0 \psi
= \frac{1}{4} \frac{a_2}{a_1} \left(\frac{\dot
a_1}{a_1} - \frac{\dot a_2}{a_2}\right) A^3, \label{Ttot12bi}\\
T_3^2 & = & -\frac{\imath}{4} \frac{a_3}{a_2} \left(\frac{\dot
a_2}{a_2} - \frac{\dot a_3}{a_3}\right)  \bp \bg^2 \bg^3 \bg^0 \psi
= \frac{1}{4} \frac{a_3}{a_2} \left(\frac{\dot
a_2}{a_2} - \frac{\dot a_3}{a_3}\right) A^1, \label{Ttot23bi}\\
T_3^1 & = &  -\frac{\imath}{4} \frac{a_3}{a_1} \left(\frac{\dot
a_3}{a_3} - \frac{\dot a_1}{a_1}\right)  \bp \bg^3 \bg^1 \bg^0 \psi
= \frac{1}{4} \frac{a_3}{a_1} \left(\frac{\dot a_3}{a_3} -
\frac{\dot a_1}{a_1}\right)A^2, \label{Ttot31bi}
\end{eqnarray}
\end{subequations}
where $A^\mu = \bp \gamma^5 \gamma^\mu \psi$ is the pseudovector.

Taking into account that in our case,  $\Box S = \ddot S +
\frac{\dot V}{V} \dot S$, in view of  \eqref{nablaS} and
\eqref{Ttotbi} for the metric \eqref{bi} from \eqref{EE} we find
\begin{subequations}
\label{EEComp}
\begin{eqnarray}
\frac{\ddot a_2}{a_2} +\frac{\ddot a_3}{a_3} + \frac{\dot
a_2}{a_2}\frac{\dot a_3}{a_3}&=& \frac{1}{\left(1 + \lambda_1 S
\right)} \left[\lambda \left( F(S) - S F_S\right) - \lambda_1 \ddot
S - \lambda_1 \left(\frac{\dot a_2}{a_2} +
\frac{\dot a_3}{a_3}\right)\dot S\right],\label{11bi}\\
\frac{\ddot a_3}{a_3} +\frac{\ddot a_1}{a_1} + \frac{\dot
a_3}{a_3}\frac{\dot a_1}{a_1}&=& \frac{1}{\left(1 + \lambda_1 S
\right)} \left[\lambda \left( F(S) - S F_S\right) - \lambda_1 \ddot
S - \lambda_1 \left(\frac{\dot a_3}{a_3} +
\frac{\dot a_1}{a_1}\right)\dot S\right],\label{22bi}\\
\frac{\ddot a_1}{a_1} +\frac{\ddot a_2}{a_2} + \frac{\dot
a_1}{a_1}\frac{\dot a_2}{a_2}&=&  \frac{1}{\left(1 + \lambda_1 S
\right)} \left[\lambda \left( F(S) - S F_S\right) - \lambda_1 \ddot
S - \lambda_1 \left(\frac{\dot a_1}{a_1} +
\frac{\dot a_2}{a_2}\right)\dot S\right],\label{33bi}\\
\frac{\dot a_1}{a_1}\frac{\dot a_2}{a_2} +\frac{\dot
a_2}{a_2}\frac{\dot a_3}{a_3} +\frac{\dot a_3}{a_3}\frac{\dot
a_1}{a_1}&=& \frac{1}{\left(1 + \lambda_1 S
\right)} \left[ \left(m S + \lambda F(S)\right) - \lambda_1 \frac{\dot V}{V} \dot S\right], \label{00bi}\\
0 &=& \left(\frac{\dot
a_1}{a_1} - \frac{\dot a_2}{a_2}\right) A^3,\label{12acbi}\\
0 &=&  \left(\frac{\dot
a_2}{a_2} - \frac{\dot a_3}{a_3}\right)  A^2,\label{23acbi}\\
0 &=& \left(\frac{\dot a_3}{a_3} - \frac{\dot a_1}{a_1}\right)
A^1.\label{31acbi}
\end{eqnarray}
\end{subequations}

From the equations \eqref{12acbi}, \eqref{23acbi} and \eqref{31acbi}
we find there exist three possibilities.

{\bf (i)} Imposing the restrictions on the spinor field only we get
\begin{equation} A^3 = A^2 = A^1 = 0. \label{spinor123bi}
\end{equation}
In this case $a_1 \ne a_2 \ne a_3$ that is the space-time
corresponds to a general Bianchi type-I model.

{\bf (ii)} By imposing restrictions on both metric functions and
spinor field we find say
\begin{equation}
\frac{\dot a_2}{a_2} - \frac{\dot a_3}{a_3} = 0, \label{A23bi}
\end{equation}
together with
\begin{equation}
A^2 = A^3 = 0. \label{spinor523bi}
\end{equation}

From \eqref{A23bi} we find

\begin{equation}
a_2 = c_1 a_3, \quad c_1 = {\rm const.} \label{a23=}
\end{equation}
Upon inserting \eqref{a23=} into \eqref{bi} the general Bianchi
type-$I$ space-time transforms into a locally rotationally symmetric
(LRS) Bianchi type-$I$ space-time.

{\bf (iii)}Finally imposing the restriction completely on the metric
functions only from \eqref{12acbi}, \eqref{23acbi} and
\eqref{31acbi} we find
 \begin{equation}
\frac{\dot a_1}{a_1} - \frac{\dot a_2}{a_2} =  \frac{\dot a_2}{a_2}
- \frac{\dot a_3}{a_3} = \frac{\dot a_3}{a_3} - \frac{\dot a_1}{a_1}
= 0, \label{dota1230bi}
\end{equation}
which can be rewritten as
\begin{equation}
\frac{\dot a_1}{a_1} = \frac{\dot a_2}{a_2} = \frac{\dot a_3}{a_3}
\equiv \frac{\dot a}{a}. \label{dota123bi}
\end{equation}
Thus in this case the Bianchi type-$I$space-time transforms into an
isotropic and homogeneous Friedmann-Robertson-Walker ($FRW$)
space-time. In what follows we study these three cases in details.

{\bf Case I} Let us recall that $A^\mu = \{A^0,\,A^1,\,A^2,\,A^3\} =
\bp  \gamma^5 \gamma^\mu \psi$ is the pseudovector. We can construct
a vector $v^\mu = \{v^0,\,v^1,\,v^2,\,v^3\}= \bp \gamma^\mu \psi.$
In view of \eqref{spinor123bi} from the equality
\begin{equation}
A_\mu v^\mu = 0, \label{Avbi}
\end{equation}
we find
\begin{equation}
A_0 v^0 = \bp  \gamma^5 \gamma_0 \psi \bp \gamma^0 \psi = \bp
\gamma^5 \gamma^0 \psi \bp^* \psi = 0. \label{Av0bi}
\end{equation}
Since $ \bp^* \psi \ne 0$, from \eqref{Av0bi} follows that $A^0 =
0$, hence $I_A = A_\mu A^\mu = 0$. But according to the Fierz
identity $I_v = - I_A = I + J$ and $I_T = I - J.$ Hence we obtain
\begin{equation}
I_A = - (S^2 + P^2) =  0, \label{AvSPbi}
\end{equation}
which leads to the fact that
\begin{equation}
S = \bp \psi = 0, \quad P = i \bp \gamma^5 \psi = 0. \label{SP0bi}
\end{equation}
Thus we conclude that if the restriction is imposed only on the
spinor field, it becomes linear and massless. Moreover, the system
becomes minimally coupled, since the coupling term $\lambda_1 R S$
vanishes. The diagonal components of Einstein equations takes the
form
\begin{subequations}
\label{EECompbig}
\begin{eqnarray}
\frac{\ddot a_2}{a_2} +\frac{\ddot a_3}{a_3} + \frac{\dot
a_2}{a_2}\frac{\dot a_3}{a_3}&=& 0,\label{11big}\\
\frac{\ddot a_3}{a_3} +\frac{\ddot a_1}{a_1} + \frac{\dot
a_3}{a_3}\frac{\dot a_1}{a_1}&=&0,\label{22big}\\
\frac{\ddot a_1}{a_1} +\frac{\ddot a_2}{a_2} + \frac{\dot
a_1}{a_1}\frac{\dot a_2}{a_2}&=&  0,\label{33big}\\
\frac{\dot a_1}{a_1}\frac{\dot a_2}{a_2} +\frac{\dot
a_2}{a_2}\frac{\dot a_3}{a_3} +\frac{\dot a_3}{a_3}\frac{\dot
a_1}{a_1}&=& 0. \label{00big}
\end{eqnarray}
\end{subequations}
As one sees, in this case the system correspond to the vacuum
solution of Einstein equation. The left hand side of
\eqref{EECompbig} can be rearranged that gives the equation for
volume scale $V$:
\begin{eqnarray}
\frac{\ddot V}{V} = 0, \label{detvlinmbi}
\end{eqnarray}
with the solution
\begin{equation}
V = V_1 t + V_2, \quad V_1, V_2 - {\rm consts.} \label{V-0}
\end{equation}
Thus we see, in this case volume scale is a linear function of $t$.
For the metric functions we obtain
\begin{equation}
a_i = D_i \left(V_1 t + V_2\right)^{\frac{1}{3} + \frac{X_i}{V_1}},
\quad \prod_{i=1}^{3} D_i = 1, \quad \sum_{i=1}^3 X_i = 0.
\label{ai-0}
\end{equation}
In this case $\frac{a_i}{a}\Bigl|_{t \to \infty} = \left(V_1 t +
V_2\right)^{X_i/V_1}\Bigl|_{t \to \infty} \nrightarrow {\rm const.}$
It means in absence of nonlinearity no isotropization takes place.

{\bf Case II}

In this case we have LRS Bianchi type-I cosmological model with $a_2
= c_1 a_3$. In this case the diagonal components of Einstein
equations can be rewritten as
\begin{subequations}
\label{EECompLRSBI}
\begin{eqnarray}
2\frac{\ddot a_2}{a_2} + \left(\frac{\dot a_2}{a_2}\right)^2&=&
\frac{1}{\left(1 + \lambda_1 S \right)} \left[\lambda \left( F(S) -
S F_S\right) - \lambda_1 \ddot S - 2 \lambda_1\frac{\dot a_2}{a_2}
\dot S\right],\label{11biLRSBI}\\
\frac{\ddot a_1}{a_1} +\frac{\ddot a_2}{a_2} + \frac{\dot
a_1}{a_1}\frac{\dot a_2}{a_2}&=&  \frac{1}{\left(1 + \lambda_1 S
\right)} \left[\lambda \left( F(S) - S F_S\right) - \lambda_1 \ddot
S - \lambda_1 \left(\frac{\dot a_1}{a_1} +
\frac{\dot a_2}{a_2}\right)\dot S\right],\label{33biLRSBI}\\
2\frac{\dot a_1}{a_1}\frac{\dot a_2}{a_2} +\left(\frac{\dot
a_2}{a_2}\right)^2 &=& \frac{1}{\left(1 + \lambda_1 S
\right)} \left[ \left(m S + \lambda F(S)\right) - \lambda_1 \frac{\dot V}{V} \dot S\right]. \label{00biLRSBI}\\
\end{eqnarray}
\end{subequations}

Subtraction of \eqref{33biLRSBI} from \eqref{11biLRSBI} gives
\begin{equation}
\frac{\ddot a_2}{a_2} - \frac{\ddot a_1}{a_1} + \frac{\dot
a_2}{a_2}\left(\frac{\dot a_2}{a_2} - \frac{\dot a_1}{a_1}\right) =
- \frac{\lambda_1}{1 + \lambda_1 S} \left(\frac{\dot a_2}{a_2} -
\frac{\dot a_1}{a_1}\right) \dot S. \label{sub21}
\end{equation}
Taking into account that $V = a_1 a_2^2$ \eqref{sub21} can be
rewritten as

\begin{equation}
\frac{d}{dt}\left(\frac{\dot a_2}{a_2} - \frac{\dot a_1}{a_1}\right)
= -\left(\frac{\dot a_2}{a_2} - \frac{\dot a_1}{a_1}\right) \frac{V
+ 2 \lambda_1 C_0}{V + \lambda_1 C_0}\frac{\dot V}{V}.
\label{sub21n}
\end{equation}

The foregoing equation can be integrated to obtain
\begin{equation}
\frac{\dot a_2}{a_2} - \frac{\dot a_1}{a_1} = \frac{X \left(V +
\lambda_1 C_0\right)}{V^2}, \quad X = {\rm const.} \label{1stInt}
\end{equation}
which gives
\begin{equation}
a_2 = a_1 \exp{\left[X \int \frac{\left(V + \lambda_1
C_0\right)}{V^2} dt \right]}. \label{a2a1}
\end{equation}
Finally on account of $a_1 a_2^2 = V$ for the metric functions we
finally obtain
\begin{equation}
a_1 = V^{1/3} \exp{\left[- \frac{2X}{3} \int \frac{\left(V +
\lambda_1 C_0\right)}{V^2} dt \right]}, \quad a_2 = V^{1/3}
\exp{\left[\frac{X}{3} \int \frac{\left(V + \lambda_1
C_0\right)}{V^2} dt \right]}. \label{a2a1final}
\end{equation}
As one sees, in case of minimal coupling, i.e. for $\lambda_1 = 0$
coincides with the results obtained in earlier papers.

\begin{figure}[ht]
\centering
\includegraphics[height=70mm]{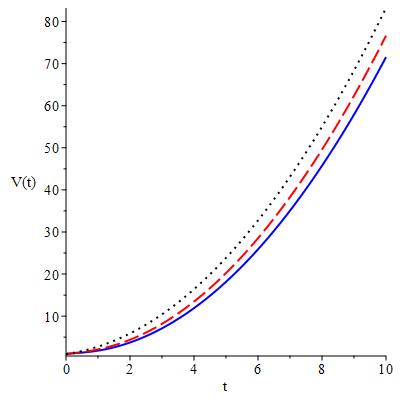} \\ \vskip 1 cm
\caption{Plot of volume scale $V$ for three different cases with $n
= 4.$.} \label{3in1n=4}
\end{figure}

Thus the metric functions are now expressed in terms of V. For the
volume scale $V$ from \eqref{EECompLRSBI} we find
\begin{equation}
\ddot V = \frac{1}{2\left(1 + \lambda_1 S\right)} \left[3 m S + 3
\lambda \left(2 F - S F_S\right) - 3 \lambda_1 \ddot S - 5 \lambda_1
\frac{\dot V}{V} \dot S\right] V. \label{V}
\end{equation}
Further taking into account \eqref{SV} we find

\begin{equation}
\ddot V = \frac{V}{\left(2V - \lambda_1 C_0\right)}\left[3 m C_0 + 3
\lambda \left(2 F - S F_S\right) - \lambda_1 C_0  \left(\frac{\dot
V}{V}\right)^2 \right]. \label{V1}
\end{equation}
If we consider the spinor field nonlinearity be a power law, say $F
= S^n$ then on account of \eqref{SV} we find
\begin{equation}
\ddot V = \frac{V}{\left(2V - \lambda_1 C_0\right)}\left[3 m C_0 + 3
\lambda \left(2 - n\right) \frac{C_0^n}{V^n}   - \lambda_1 C_0
\left(\frac{\dot V}{V}\right)^2 \right]. \label{V2}
\end{equation}
We solve this equation numerically. For simplicity we set $m = 1$
and $C_0 = 1$. We consider three case setting $\lambda_1 = 1,\,
\lambda = 1$ (non-minimal coupling with nonlinear term, blue solid
line), $\lambda_1 = 0,\, \lambda = 1$ (minimal coupling with
nonlinear term, red dash line) and $\lambda_1 = 1,\, \lambda = 0$
(non-minimal coupling without nonlinear term, black dot line). In
case of nonlinear spinor field we set $n = 4$. As the initial
condition we set $V(0) = 1$ and $\dot V(0) = 1$. The evolution of
the volume scale $V(t)$ is given in Fig. \ref{3in1n=4}

\begin{figure}[ht]
\centering
\includegraphics[height=70mm]{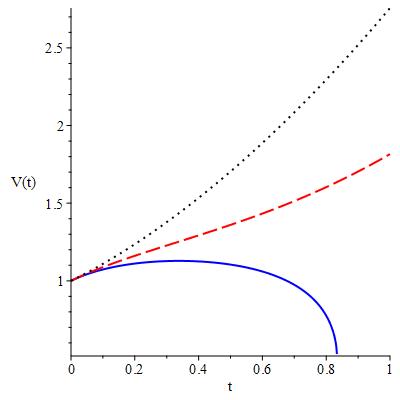} \\ \vskip 1 cm
\caption{Plot of volume scale $V$ for three different cases with $n
= 5.$.} \label{3in1n=5}
\end{figure}

In Fig. \ref{3in1n=5} we have plotted the evolution of the Universe
as in previous case only with $n = 5$. In this case for non-minimal
coupling with spinor field nonlinearity we see the Universe is
closed. After attaining some maximum value the Universe begins to
shrink and ends in Big Crunch. It should be noted that  in our
earlier study with minimal coupling no such results were obtained.

As far as FRW case is concerned, we will study this model in some
forthcoming paper.

\section{Discussion and conclusion}

Here let us point out a few things. As we have already mentions the
spinor field is very sensitive to gravitational one one and the
covariant derivative acts on spinor field in a definite way, namely
\begin{equation}
\nabla_\mu \psi = \partial_\mu \psi - \Gamma_\mu \psi, \quad
\nabla_\mu \bp = \partial_\mu \bp + \bp \Gamma_\mu. \nonumber
\end{equation}
While working with non-minimal coupling we have some construction
like $\nabla_\mu \nabla_\nu S$, where $S = \bp \psi$ is a scalar. In
this paper we used the property od the spinor field that gives
$\nabla_\mu S = \partial_\mu S$. But what if we use the spinor
notation? In that case we have
\begin{eqnarray}
\nabla_\mu S &=& \nabla_\mu \left(\bp \psi\right) = \left(\nabla_\mu
\bp \right) \psi + \bp \left(\nabla_\mu \psi\right) \nonumber\\
&=& \left(\partial_\mu \bp + \bp \Gamma_\mu\right) \psi + \bp
\left(\partial_\mu \psi - \Gamma_\mu \psi\right) =
\left(\partial_\mu \bp \right) \psi + \bp \left(\partial_\mu
\psi\right) = \partial_\mu \left(\bp \psi\right) = \partial_\mu S.
\nonumber
\end{eqnarray}

What happens to second derivative?

In one hand
\begin{eqnarray}
\nabla_\nu \nabla_\mu S = \nabla_\nu \partial_\mu S =
\partial_\nu \partial_\mu S - \Gamma_{\nu \mu}^\tau \partial_\tau S.
\label{ddcovS1}
\end{eqnarray}
On the other hand we have
\begin{eqnarray}
\nabla_\nu \nabla_\mu S &=& \nabla_\nu \partial_\mu S = \nabla_\nu
\left(\partial_\mu \bp \psi + \bp \partial_\mu \psi\right)
\nonumber\\
&=& \left(\nabla_\nu \partial_\mu \bp\right) \psi + \partial_\mu \bp
\nabla_\nu \psi + \nabla_\nu \bp \partial_\mu \psi + \bp
\left(\nabla_\nu \partial_\mu \psi\right) \nonumber \\
&=& \left(\partial_\nu \partial_\mu \bp - \Gamma_{\nu \mu}^\tau
\partial_\tau \bp\right) \psi + \partial_\mu \bp \left(\partial_\nu
\psi - \Gamma_\nu \psi\right) \nonumber\\ &+& \left(\partial_\nu \bp
+ \bp \Gamma_\nu \right) \partial_\mu \psi + \bp \left(\partial_\nu
\partial_\mu \psi - \Gamma_{\nu \mu}^\tau
\partial_\tau \psi\right)\nonumber\\
&=& \partial_\nu \partial_\mu S - \Gamma_{\nu\mu}^\tau \partial_\tau
S + \left(\bp \Gamma_\nu \partial_\mu \psi - \partial_\mu \bp
\Gamma_\nu \psi\right). \label{ddcovS}
\end{eqnarray}

So in order to get the both \eqref{ddcovS1} and \eqref{ddcovS}
identical, we should have
\begin{equation}
\bp \Gamma_\nu \partial_\mu \psi -
\partial_\mu \bp \Gamma_\nu \psi = 0. \label{addcon}
\end{equation}
In our case spinor field depends on $t$ only, whereas $\Gamma_0 =
0$. Taking into account that $\Gamma_i = (\dot a_i /2) \bg^i \bg^0$,
where $i= 1,\,2,\,3$ we rewrite the left hand side of \eqref{addcon}
as follows
\begin{eqnarray}
\bp \Gamma_\nu \partial_\mu \psi -
\partial_\mu \bp \Gamma_\nu \psi = \frac{\dot a_i}{2} \left(\bp
\bg^i \bg^0 \dot \psi - \dot \bp \bg^i \bg^0 \psi\right),
\label{addcon1}
\end{eqnarray}
which, on account of \eqref{speqbi} can be written as
\begin{eqnarray}
\bp \Gamma_\nu \partial_\mu \psi -
\partial_\mu \bp \Gamma_\nu \psi = - \dot a_i \frac{\dot V}{V} \bp
\bg^i \psi = - a_i \dot a_i \frac{\dot V}{V} \bp \gamma^i \psi = -
a_i \dot a_i \frac{\dot V}{V} v^i. \label{addcon2}
\end{eqnarray}
As it was shown earlier $v^\mu = \bp \gamma^\mu \psi$ is the vector,
constructed spinor fields and in case of BI cosmology it is trivial.
As far as LRS-BI or FRW models are concerned, the demand that both
\eqref{ddcovS1} and \eqref{ddcovS} are identical imposes the
following restrictions on the components of the spinor field:
\begin{eqnarray}
\psi^*_1 \psi_4 + \psi^*_2 \psi_3 + \psi^*_3 \psi_2 + \psi^*_4
\psi_1 &=& 0 \nonumber\\
\psi^*_1 \psi_4 - \psi^*_2 \psi_3 + \psi^*_3 \psi_2 - \psi^*_4
\psi_1 &=& 0 \nonumber\\
\psi^*_1 \psi_3 - \psi^*_2 \psi_4 + \psi^*_3 \psi_1 - \psi^*_4
\psi_2 &=& 0 \nonumber
\end{eqnarray}

Finally we can make the following conclusions. The consideration of
non-minimal coupling has no effect on the non-diagonal components of
the energy-momentum tensor of the spinor field. As a result, the
restrictions on the space-time geometry remain the same as in case
of minimal coupling. Nevertheless, the diagonal components of EMT
differ. As one sees, while the linear spinor field with non-minimal
coupling or non-linear spinor field with minimal coupling in some
cases give rise to open universe, the nonlinear spinor field with
non-minimal coupling with the same parameters generates model that
is close, i.e., after attaining some maximum value begins to
decrease and finally shrinks to Big Crunch.

\end{document}